\documentclass[]{interact}

\usepackage{epstopdf}
\usepackage[caption=false]{subfig}

\usepackage[numbers,sort&compress]{natbib}
\bibpunct[, ]{[}{]}{,}{n}{,}{,}
\renewcommand\bibfont{\fontsize{10}{12}\selectfont}
\makeatletter
\def\NAT@def@citea{\def\@citea{\NAT@separator}}
\makeatother

\theoremstyle{plain}

\theoremstyle{definition}

\theoremstyle{remark}

\usepackage{graphicx}
\usepackage{dcolumn}
\usepackage{bm}
\usepackage{amsmath}
\usepackage{soul}
\usepackage{color}

\begin{document}


\title{Universal and nonuniversal statistics of transmission in thin random layered media}

\author{
\name{Jongchul Park\textsuperscript{a,b}, Matthieu Davy\textsuperscript{c},Victor A. Gopar\textsuperscript{d}, and Azriel Z. Genack\textsuperscript{a}\thanks{CONTACT A.~Z.~Genack. Email: agenack@qc.cuny.edu, V.A. Gopar. Email: gopar@unizar.es} }
\affil{\textsuperscript{a}Department of Physics, Queens College and The Graduate Center of the City
University of New York, Flushing, NY, 11367 USA; \textsuperscript{b}Chiral Photonics Inc., Pine Brook, NJ 07058, USA; \textsuperscript{c}Univ de Rennes, CNRS, Institut d'électronique et des technologies du numérique (IETR) -UMR 6164, 35000 Rennes, France; \textsuperscript{d}Departamento de F\'isica Te\'orica and BIFI, Universidad de Zaragoza, Pedro Cerbuna 12, E-50009, Zaragoza, Spain}
}

\maketitle

\begin{abstract}
The statistics of transmission through random 1D media are generally presumed to be universal and to depend only upon a single dimensionless parameter—the ratio of the sample length and the mean free path, $s=L/\ell$. Here, we show in numerical simulations and optical measurements of random binary systems, and most prominently in systems for which $s < 1$, that the statistics of the logarithm of transmission, $\ln T$, are universal for transmission near the upper cutoff of unity and depend distinctively upon the reflectivity of the layer interfaces and their number near a lower cutoff. The universal segment of the probability distribution function of the logarithm of transmission $P(\ln T)$ is manifested with as few as three binary layers. For a given value of $s$, $P(\ln T)$ evolves towards a universal distribution as the number of layers increases. Optical measurements in stacks of 5 and 20 glass coverslips exhibit statistics at low and moderate values of transmission that are close to those found in simulations for 1D layered media, while differences appear at higher transmission where the transmission time in the medium is longer and the wave explores the transverse nonuniformity of the sample. 
\end{abstract}


\section{Introduction}

The scattering of optical, acoustic, and electron waves plays a dominant role in our ability to communicate and image. Wave interactions with natural and fabricated disordered materials are crucial in medical imaging, telecommunication, and resource exploration.

We will consider the nature of waves in random 1D systems, which appear in a wide range of contexts.
Measurements have been carried out for acoustic waves reflected from the stratified crust of the
Earth~\cite{Sato}, and launched along a string with randomly positioned weights~\cite{He}.  Also, light
transmitted through stacks of dielectric slabs of random thickness \cite{Liew,Moretti}, including overhead
transparencies~\cite{Berry}, glass coverslips separated by air~\cite{Zhang}, evaporated multilayer
coatings, and microporous silicon layers~\cite{Bertolotti} have been experimentally studied.

Measurements have also been carried out in single-mode
microwave waveguides~\cite{Sebbah,Antonio_FM,Cheng,Huang} and in  optical waveguides produced in channels between
periodic arrays of holes in a dielectric wafer~\cite{Topolancik,Lodahl}, in open metal systems
supporting spoof surface plasmon polaritons~\cite{Sahoo} and in Bose-Einstein condensates in random
potentials created by interfering light beams~\cite{Fallani,Billy}. Disorder can degrade the
performance of devices based on the presence of a stop band in periodic structures such as in fiber
Bragg gratings used in telecommunications~\cite{Othonos} and  semiconductor vertical cavity
surface emitting lasers~\cite{Koyama}. But disorder may also beneficially facilitate excitation deep
within a diffusive sample and can reduce the lasing threshold in amplifying random media since energy
can be deposited deep inside random structures in long-lived localized modes peaked near the center of the sample~\cite{Milner,Liu}. In an initially periodic medium,  transmission will increase for any added
disorder~\cite{Freilikher}.

In the late 1950’s,  it was demonstrated theoretically that the diffusion of waves  may cease~\cite{Anderson58}. The suppression of transport is associated with 
 constructive interference of partial waves following time-reversed trajectories returning to points in
the medium~\cite{Altshuler}. Anderson showed that beyond a threshold in disorder of the onsite energy, electrons are exponentially localized in 3D disordered lattices~\cite{Anderson58}, while Gertsenshtein and Vasil'ev showed that the
average transmission of radio waves in a single-mode waveguide decays exponentially with length for
any strength of disorder \cite{Gertsenshtein-Vasilev}.

Since transport in disordered systems is a random process on any length scale, it is important to find the 
scaling of the full probability distribution functions (pdfs) of the transmission or the conductance and of other
propagation variables, in addition to the averages of these quantities. A wide range of transport
phenomena in random 1D media have been explored
theoretically~\cite{Pendry,Azbel,Sipe,Yeh,Freilikher,Deych,Fouque,Nascimento, Gopar,Izrailev}.

The statistics of transmission were first investigated in random 1D media.
The ensemble average of the logarithm of transmission in 1D is given by $\langle -\ln T \rangle=L/\ell \equiv s$~\cite{Anderson80}. For samples with length $L$ much greater than the mean
free path $\ell$, $s \gg 1$, the pdf of transmission approaches a log-normal
distribution~\cite{Melnikov,Abrikosov} with var$(\ln T)=2s-{\pi^2}/3$ \cite{Cheng}. Propagation in
this limit is thus determined by a single parameter, $s$, and is independent of the details of scattering. 
The pdf of the electronic conductance in disordered 1D quantum wires was studied for all lengths with
the aid of random matrix theory (RMT) for samples in the dense-weak-scattering limit:$\ell \gg \lambda_F$, where $\lambda_F$ is the Fermi wavelength,
in which the
number of scatterers diverges as the scattering strength of individual elements tends to zero~\cite{Mello-book}. 
The evolution of the distribution
of transmission with system length $L$ is described by the differential equation \cite{Melnikov,
Mello-book}:
\begin{equation}
\label{1Ddiffusion}
\frac{\partial P_s(\mu)}{\partial s }=\frac{\partial}{\partial \mu}\left[\mu\left(\mu+1 \right) \frac{\partial
P_s(\mu)}{\partial \mu} \right],
\end{equation}
where $\mu=1/T-1$. 
Equation~(\ref{1Ddiffusion}) was solved subject to the ballistic initial condition: $\lim_{s \to 0 }
P_{s}(\mu)=\delta (\mu)$. The solution of Eq.~(\ref{1Ddiffusion}) \cite{Gertsenshtein-Vasilev,
Abrikosov}, gives the pdf of the logarithmic transmission
\begin{equation}
\label{poflnT_exact}
P_s(\ln T)= \frac{s^{-\frac{3}{2}}}{\sqrt{2\pi}} \frac{{\rm
e}^{-\frac{s}{4}}}{T}\int_{y_0}^{\infty}dy\frac{y{\rm
e}^{-\frac{y^2}{4s}}} {\sqrt{\cosh{y}+1-2/T}},
\end{equation}
where $y_0={\rm arcosh}{(2/T-1)}$. According to Eq.~(\ref{poflnT_exact}), wave propagation is determined by the single
parameter $s$, which is independent of the details of scattering~\cite{Mello_1988}. The result of universal wave statistics has been so important in the history of the field of wave transport in disordered media that the practical case of a small number of layers has not been explored.

In this article, we treat a common embodiment of a random 1D medium of a small number of layers
with random thickness, which naturally includes samples thinner than the mean free path. We carry out
simulations of the pdf of transmission at different lengths. Surprisingly, the form of the distributions is established in samples as thin as three binary layers.
There is a crossover within the distribution $P(\ln T)$ from 
the shape of Eq.~(\ref{poflnT_exact}) predicted by random matrix theory
in the dense-weak-scattering limit, which closely resembles a segment of a Gaussian distribution with
an upper cutoff at $\ln T=0$ ($T=1$) at higher levels of transmission to a function that depends on the
characteristics of the medium and drops sharply towards the lower cutoff in transmission. $P(\ln
T)$ just above the lower cutoff exhibits a distinctive dependence upon the number of layers and the 
surface reflectivity of the elements. We compare these results to measurements of optical transmission through stacks of glass coverslips. Though the thickness of the coverslips is not uniform, and the transmitted intensity varies strongly across the sample, so that the sample is not one-dimensional, the results 
are in good agreement with simulations for small and moderate values of transmission. 
\begin{figure}
\includegraphics[width=\columnwidth]{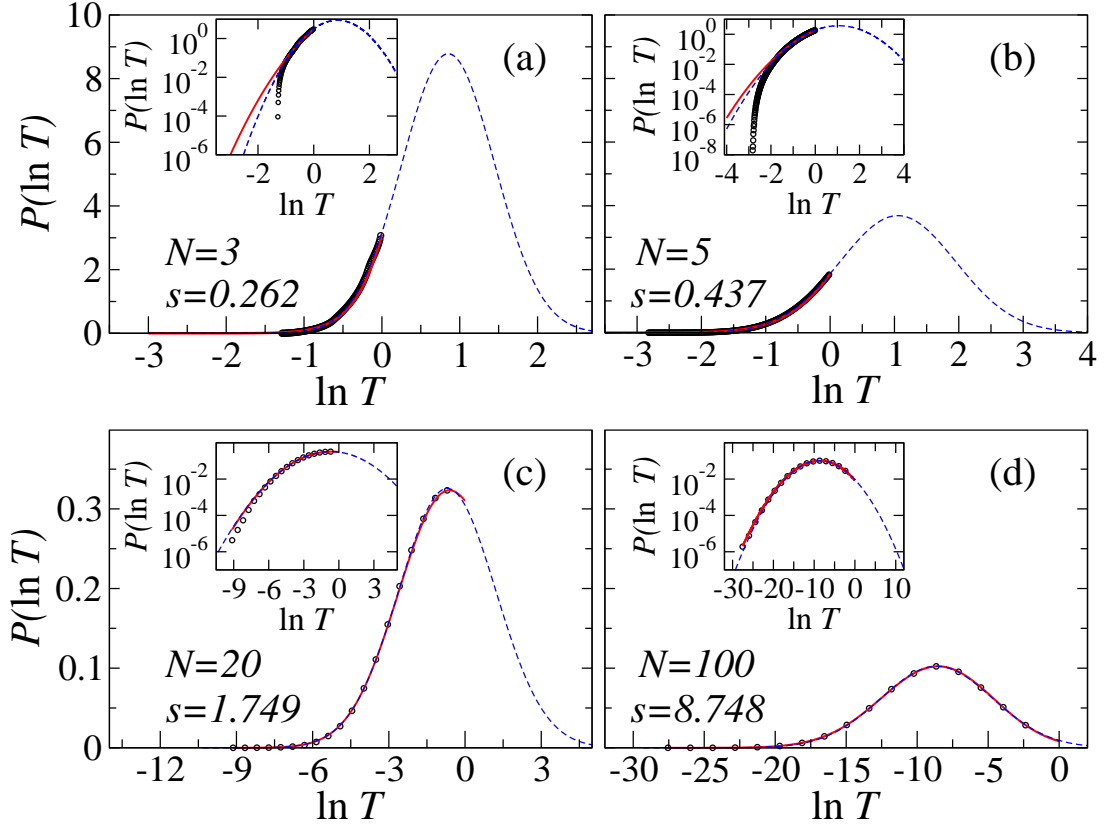}
\caption{PDFs of the logarithm of transmission $P(\ln T)$ for ensembles with
different numbers of layers. Black circles represent the numerical simulations. The red lines represent the results of Eq.~(\ref{poflnT_exact}) with parameter $s=L/l = 0.262, 0.437, 1.749$ and 8.748 with
$n = 1.5217$ for panels (a),(b), (c), and (d), respectively. The dashed blue lines are Gaussian fits of the
numerical simulations. Insets show the pdfs $P(\ln T)$ in logarithmic scale. A sharp drop near
$T_m$ is seen in panels (a) and (b). The size of the ensembles in each frame is:
$10^9$ for $N=3$, $8.73 \times 10^{10}$ for $N=5$, $8.7 \times 10^7$ for $N=20$, and  $7.3 \times 10^7$ for $N=100$.
}
\label{fig_1}
\end{figure}

\section{Numerical simulations of wave transmission}

We carry out scattering matrix simulations of wave propagation through a fixed number of binary
layers. This corresponds to an optical plane wave
incident normally upon alternating uniform parallel dielectric layers with air spaces with random thickness of dielectric and air layers. The randomness in
the thickness of each layer with refractive indices of $n$ and 1 is much greater than the wavelength, so there is no correlation in the phase of the electromagnetic field at adjacent interfaces. This is
accomplished in simulations in which the vacuum wavelength is scanned between 600 and 700 nm
and the thicknesses of each of the layers is drawn from a normal
distribution with an average thickness of $50$ $\mu$m and a standard deviation of $5$  $\mu$m. Since all
interfaces are between media with indices 1 and $n$, the reflection coefficient is the same at each
interface. The size of the system is given by the number of dielectric layers, $N$, of index of refraction
$n$. The statistics of propagation are found for ensembles with fixed $N$ in contrast to studies
in continuous random media in which the physical length of the sample is fixed.

Results of simulations of $P(\ln T)$ for ensembles with $n=1.5217$ and different $N$ are shown in
Fig.~\ref{fig_1} (dots). The index of refraction is that of the glass coverslips used in measurements.
Two cutoffs are observed. An upper cutoff in $P(\ln T)$ at $\ln T=0$, corresponding to perfect
transmission, and a lower cutoff, which can be seen in the logarithmic plots in the insets in each of the
frames of Fig.~\ref{fig_1}. The theoretical prediction for $P(\ln T)$ given by
Eq.~(\ref{poflnT_exact}), is plotted in each frame of Fig.~\ref{fig_1} (red lines) and coincides with
the simulated results at high values of transmission, but theory and
simulations diverge for small values of transmission, as can be seen in the logarithmic plots in the insets of Figs. \ref{fig_1}(a) and~\ref{fig_1}(b). The system
under study has a clear minimum when 
the thickness of each layer corresponds to a quarter wavelength modulo a half wavelength~\cite{Freilikher,Fowles}. The layer thickness is then equal to $d_i = ( m_i +1/2 ) \lambda / 2 n_i$, where $n_i$ and $d_i$ are the indices of
refraction and thicknesses of the layers, respectively, and $m_i$ is a positive integer. This corresponds to a quarter wave stack. The minimum transmission is given by~\cite{Fowles}
\begin{equation}
\label{Tm}
T_m= 1 - \left(\frac{n^{2N}-1}{n^{2N}+1} \right)^2 .
\end{equation}

\begin{figure}
\includegraphics[width=\columnwidth]{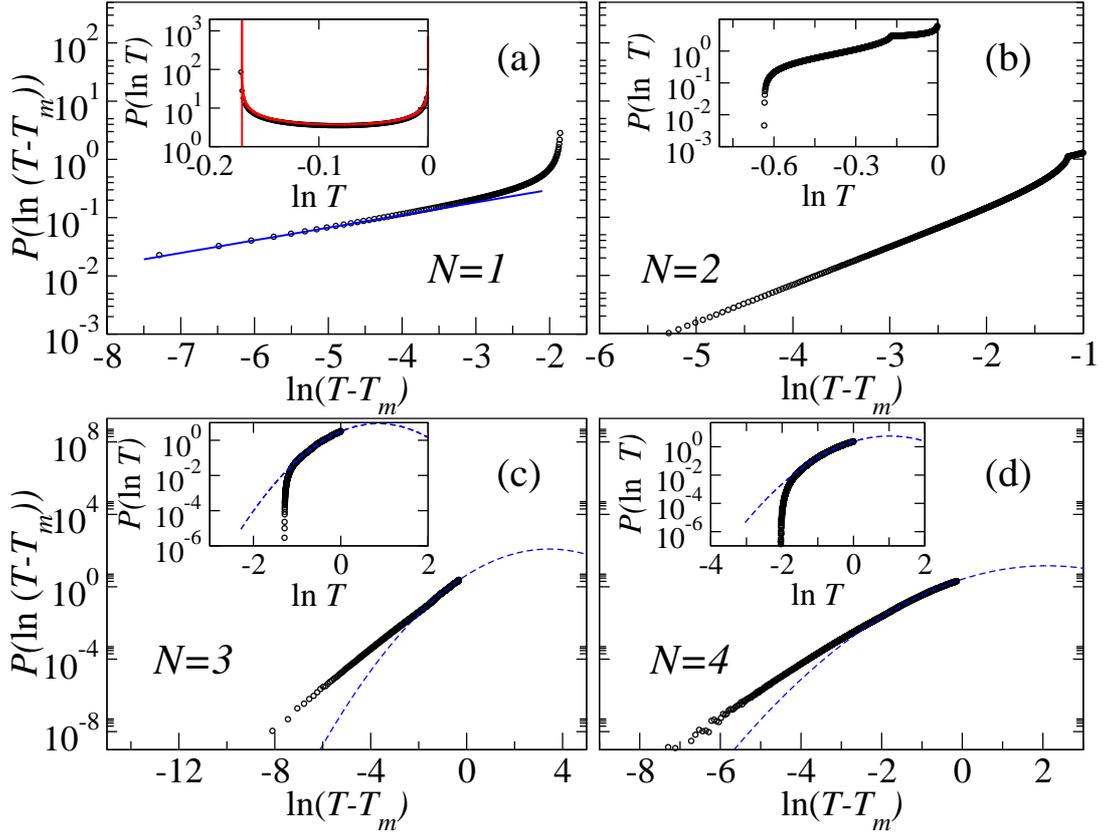}
\caption{Main frames: The pdf of the logarithm of the difference between the transmission and its
minimum value $T_m$ for different numbers of layers. Dots represent the numerical simulations, while
the dashed blue line are Gaussian fits. The insets show $P(\ln T)$ on a logarithmic scale. The red line
in inset (a) represents the analytical result obtained in Appendix A, Eq.~(\ref{poflnT_1}). The number of realizations of the disorder in each case is:
$2.1 \times 10^8$ for $N=1$, $4 \times 10^8$ for $N=2$, $10^9$ 
for $N=3$, and $1.51 \times 10^9$ for $N=4$.}
\label{fig_2_sub}
\end{figure}

For higher values of $T$, the pdfs of $\ln T$ shown in Fig.~\ref{fig_1} closely resemble segments of a
Gaussian distribution. Though $T$ cannot exceed unity, the Gaussian extending
above $\ln T =0 $ is drawn in Fig.~\ref{fig_1} (blue dashed line). The upper cutoff of the distribution
at $(\ln T=0)$ is discontinuous, while the lower cutoff is continuous above $T_m$, as seen in
the insets of Fig.~\ref{fig_1}.

\begin{figure} 
\centering
\includegraphics[width=0.7\columnwidth]{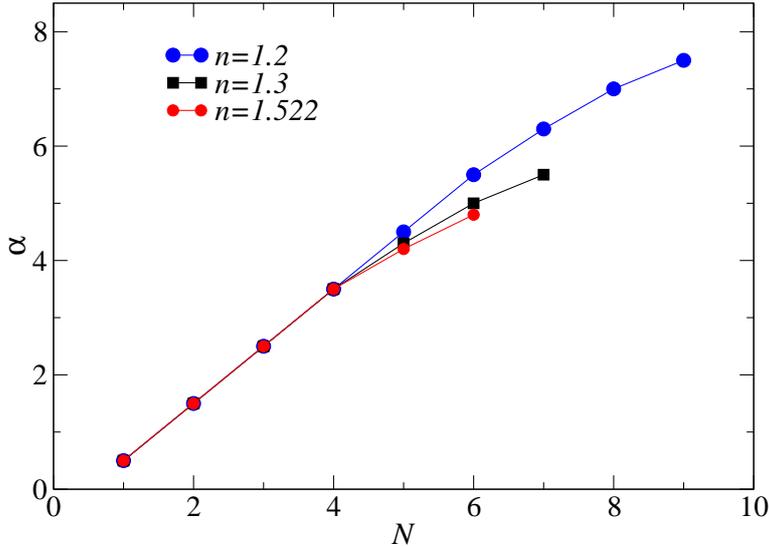}
\caption{
Exponent $\alpha$ corresponding to the slope of $\ln[P(\ln(T-T_m))]$, as explained in the text, for
different values of the index of refraction $n$ and the number of layers, $N$.}
\label{fig_3_sub}
\end{figure}

We now consider $P(\ln T)$ near $T_m$ revealed in simulations of the pdf of the difference between
the transmission and the minimum transmission.  As shown in Fig.~\ref{fig_2_sub}, $P\left( \ln ( T -T_m ) \right)$ for
$N=1, 2 ,3 , 4$, just above $T_m$ is seen to increase linearly on a logarithmic scale, so that 
\begin{equation}
P \left( \ln ( T-T_m ) \right) \propto \exp
\left[{ \alpha \ln (T - T_m ) }\right] = ( T - T_m )^\alpha.
\label{proptoalpha}
\end{equation}
The power $\alpha$ in Eq.~(\ref{proptoalpha}) is seen in Fig.~\ref{fig_3_sub} to increase linearly with $N$ for $N \le 4$ as 
\begin{equation}
\label{alpha}
\alpha=\frac{1}{2}\left(2 N-1 \right) ,
\end{equation}
and appears to be independent of $n$. For $N > 4$,  however, $\alpha$ is seen to depend on $n$,  for the values of $n$ used in the simulations.  The value $\alpha=1/2$ for a single layer, $N= 1$, which is an etalon or Fabry-Perot interferometer is derived in Appendix A.

\begin{figure}
\centering
\includegraphics[width=0.7\columnwidth]{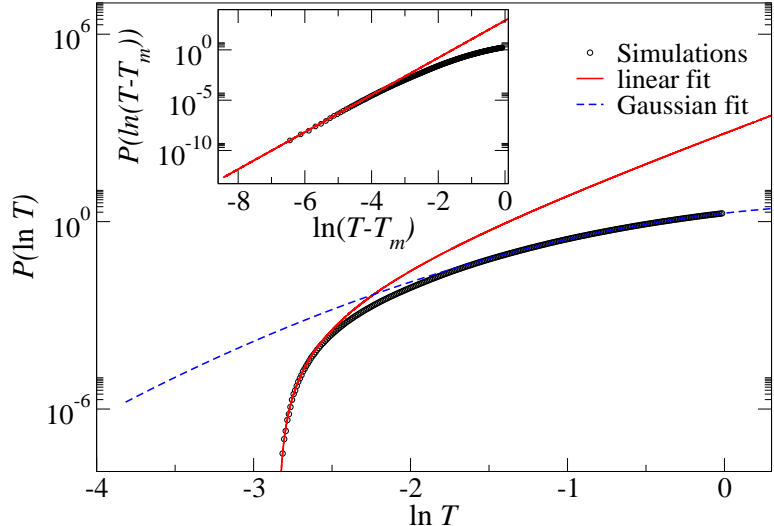}
\caption{Numerical distribution $P(\ln T)$ for $N=5$ (dots). Linear and Gaussian fits are represented by the
red and dashed-blue lines, respectively. The inset shows the linear behavior of $P(\ln (T-T_m))$ on a logarithmic scale. $P(\ln T)$ is obtained from an  ensemble of $8.73 \times 10^{10}$ random 
realizations.}
\label{fig_4_sub}
\end{figure}

Simulations for $N=5$ presented in Fig.~\ref{fig_4_sub} show that there is a crossover in the behavior of $P(\ln T)$. Above the point at which the dashed and red lines in Fig.~\ref{fig_4_sub} cross, $P(\ln T)$ closely follows a segment of a Gaussian distribution, which corresponds to the bulk of the pdf of Eq.~(\ref{poflnT_exact}), as discussed previously. 
Below the crossover, $P(\ln T)$ is determined according to Eq. (\ref{proptoalpha}) with an exponent $\alpha$ given by Eq.~(\ref{alpha}) for $N \leq 4$, and $T_m$ given by Eq. (\ref{Tm}), which is shown as the red curve.  In the crossover
regime, simulations of $P(\ln T)$ fall slightly below both curves. 

\begin{figure}
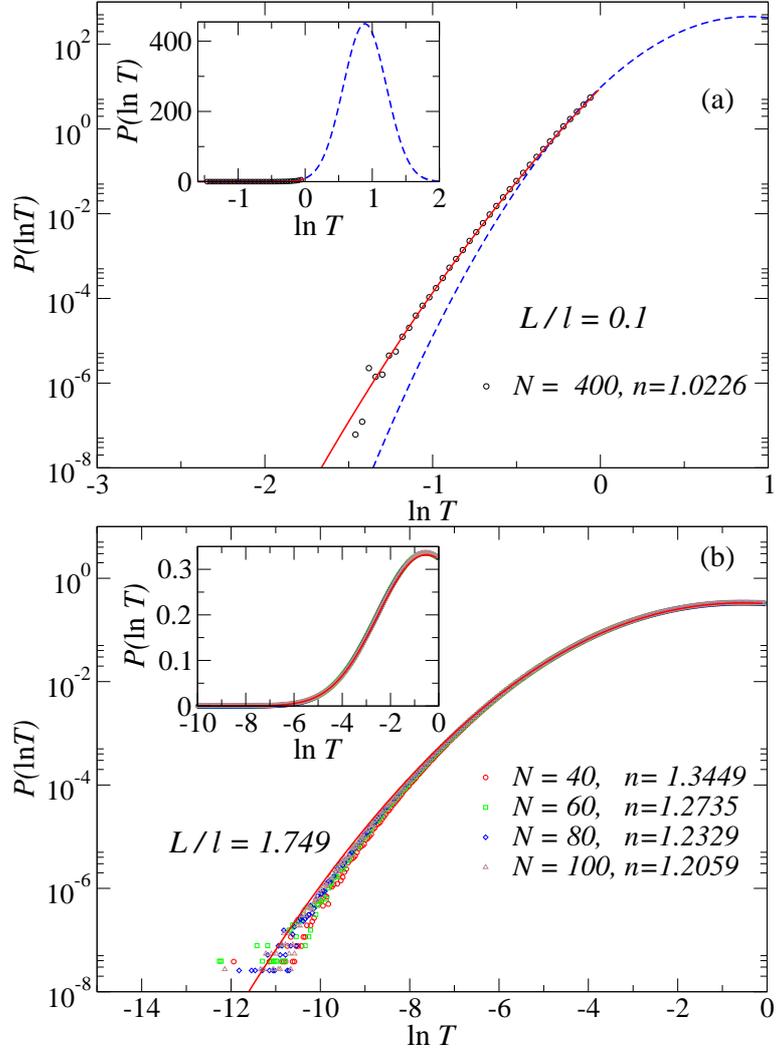

\centering
\includegraphics[width=0.7\columnwidth]{Fig_5a_RE_resub.eps}
\includegraphics[width=0.7\columnwidth]{Fig_5b_RE_resub.eps}
\caption{(a) Numerical distribution $P(\ln T)$ for a large number of layers ($N=400$) but $s=0.1$.
The red line corresponds to RMT predictions, Eq.~(\ref{poflnT_exact}), while the blue-dashed line
represents a Gaussian fit. (b) $P(\ln T)$ for different number of layers. The numerical simulations approach the predictions of RMT as the number of layers increases
(dots).}
\label{fig_5_sub}
\end{figure}

The results of simulations for a small value of $s$ but large number of layers: $N=400$ with a small index of refraction of $n=1.0226$ is shown in Fig.~\ref{fig_5_sub}(a). In this
case, $T_m (\sim 10^{-8})$ is exceedingly small and the
RMT result of Eq.~(\ref{poflnT_exact}) matches the numerical simulations, while  the Gaussian distribution  falls faster than the numerical simulations in the tail.

As the number of layers increases, $T_m$ decreases and the pdf of $\ln T$ for binary structures
approaches the results of RMT for the dense-weak-scattering limit given by Eq.~(\ref{poflnT_exact}), as seen in Fig.~\ref{fig_5_sub}(b), where $P(\ln T)$ is plotted for different values of $N$, for a fixed value of the scaling parameter, $s=1.749$ 

\section{Statistics of transmission time with high or low transmission}

\begin{figure}
\centering
\includegraphics[width=0.7\columnwidth]{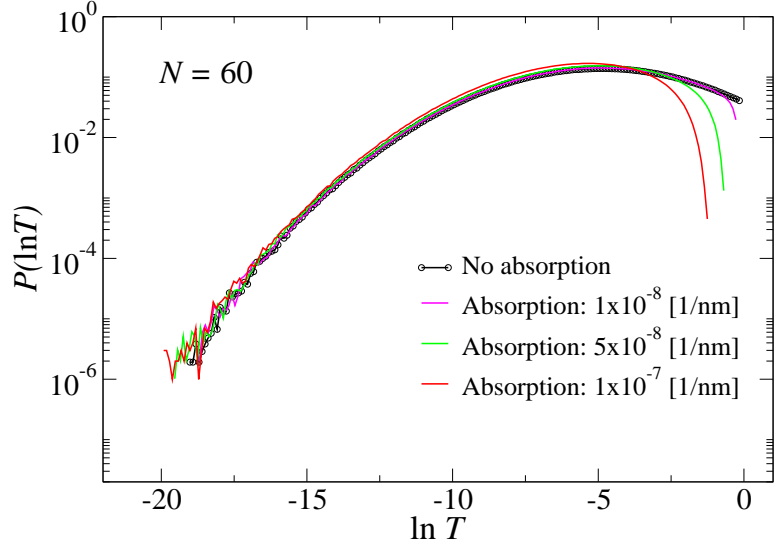}
\caption{The distribution of $\ln T$ for different values of absorption for N = 60. $P(\ln T)$ is obtained from $10^7$ different disorder realizations in each case.}
\label{fig_6_sub}
\end{figure}

Here we seek to identify the characteristics of subensembles of sample configurations with high or low transmission that might place these subensembles in the universal or the nonuniversal portions of $P(\ln T)$, respectively. To this end, we investigate the impact of absorption within the sample and the statistics of transmission time and the energy profile. We see in Fig.~\ref{fig_6_sub} that when even minimal absorption is introduced by adding a small imaginary part to the index of refraction, the transmission can no longer be perfect and the pdf of $\ln T$ at $\ln T=0$ drops to zero. As the absorption level is increased, the dip in $P(\ln T)$ moves progressively to lower values of $\ln T$ while $P(\ln T)$ is hardly affected
for very low values of transmission. The smaller impact of absorption upon waves with low transmission suggests that the transmission time is smaller in such configurations. 

The transmission time of the waves, $\tau$, is given by the spectral derivative of the phase of the transmitted field $d\phi/d\omega$ ~\cite{Wigner1955,Avishai,Genack,Texier,Davy_2015_1,Huang_2022}  and it is proportional to the density of states (DOS):
$\rho$~\cite{Davy_2015_1,Davy_2015_2}
\begin{equation}
\tau=\pi \rho
\end{equation}

The transmission time  is small when the thicknesses of all layers are close to an integer multiple of a half wavelength plus a quarter wavelength, for which the transmission is near the minimum value that occurs in the center of the first band gap. 

\begin{figure}
\centering
\includegraphics[width=0.7\columnwidth]{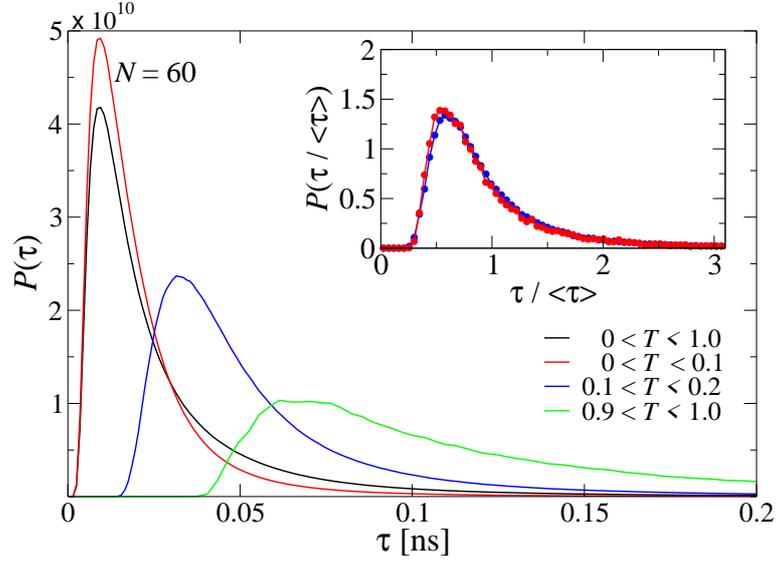}
\caption{
Distributions of transmission time  for various ranges of transmission in a medium composed of N = 60 layers: all $T$ (black), $T < 0.1$ (red), $0.1 < T < 0.2$ (blue), and $0.9 < T < 1$ (green) with $s =5.249$. The inset shows the distribution of transmission  time normalized by the average transmission time for transmission intervals, $T/<T> = 0.1 \pm 0.0001$ (blue), and $1.0 \pm 0.0001$ (red). The distributions are obtained from $10^9$ different disorder realizations.}
\label{fig_7_sub}
\end{figure}

The pdfs of the transmission time  over all samples and over subensembles with different ranges of transmission in a sample composed of 60 layers are shown in Fig.~\ref{fig_7_sub}. Both the transmission time, at which $P(\tau)$ begins to rise, and the average transmission time increases as the average transmission of the range covered increases. Surprisingly when the transmission time is normalized by its ensemble average for each transmission range, the resulting distributions overlap, as seen in the inset of Fig.~\ref{fig_7_sub}. 

\begin{figure}
\centering
\includegraphics[width=0.7\columnwidth]{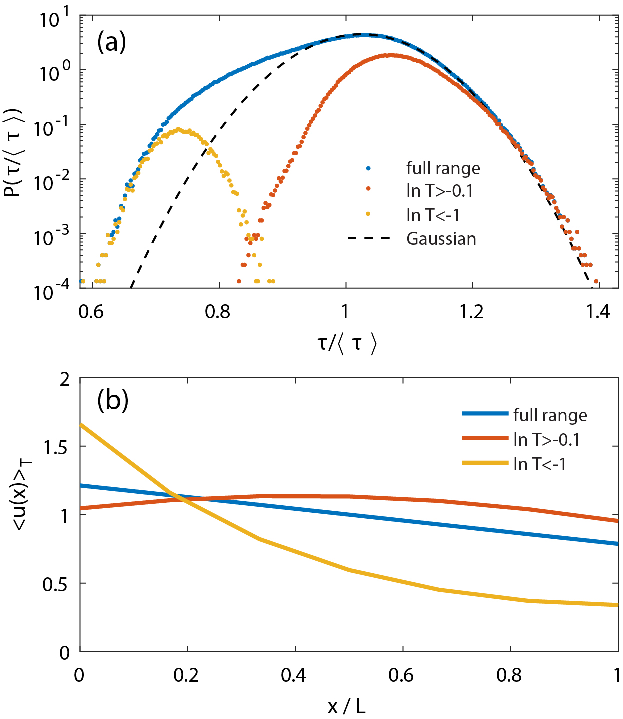}
\caption{(a) Distribution of the transmission time for samples $N=5$ dielectric layers with $\ln T > 0.1$, $\ln T < -1$, and the full range of $T$. The dashed line is a Gaussian fit. (b) Profile of average energy density for samples with the same ranges of transmission as in (a). The energy density in a single configuration is given by $\epsilon E^2/2$, where $E$ is the electromagnetic field.}
\label{fig_8_sub}
\end{figure}

The pdfs of the transmission time  for all disordered configurations with $N=5$, and for configurations with transmission near the upper and lower cutoffs, which fall in the universal and nonuniversal ranges of transmission of $P(\ln T)$, respectively, are shown in Fig.~\ref{fig_8_sub}(a). The crossover between universal and nonuniversal behavior of $P(\ln T)$ for this ensemble is near  $\ln T=-1$, as seen in Fig.~\ref{fig_1}(a). In Fig.~\ref{fig_8_sub}(a), the probability distribution $P(\tau/\langle \tau \rangle)$ for a subensemble of $T$ is normalized by the ratio of the number of configurations for the respective subensemble to the total number of configurations. $P(\tau/\langle \tau \rangle)$ for samples with high and low transmission ranges overlap the long and short time tails for the distribution for all configurations, respectively. 
This confirms that subsets with low and high transmissions correspond to configurations with short and long transmission times, respectively.

The difference in transmission times  is also naturally reflected in the intensity profiles within the sample  shown in Fig.~\ref{fig_8_sub}(b). In  dissipationless structures, the DOS is proportional to the sum of the energy density excited inside the sample by unit flux incident from both sides of the sample 
\begin{equation}
 \rho=U/2\pi ,
\end{equation}
where $U=u+u'$, and $u$ and $u'$ are the energies excited from the left and right, respectively \cite{Huang_2022}. Since the ensemble average of the energy excited within the sample is the same for excitation from the left or right, 
\begin{equation}
 \langle \tau \rangle = \langle u \rangle.
\end{equation}
This relationship holds for any subensemble as well . Since the energy falls towards the output of the samples with low transmission, the energy excited within the sample and the transmission time are relatively small.

The profiles of average energy density $\langle u(x) \rangle_T$ 
averaged over samples within a specific range of transmission $T$ 
at position $x$ inside the sample 
for unit incident flux is shown in Fig.~\ref{fig_8_sub}(b)  for $\ln T < - 1$  and for $\ln T > -0.1$. The energy density falls from the input in samples with lower transmission. In contrast, the energy density for high transmission rises to a peak near the center of the sample, as is seen in slabs \cite{Choi, Razo_2022} and quasi-1D samples \cite{Davy_2015_1,Cao,Shi}. The large integral over the sample length of energy density for configurations with higher transmission corresponds to  longer times delay, as shown in Fig.~\ref{fig_8_sub}(a). The shortest transmission  times arise in samples  that are close to corresponding to a quarter-wave stack.  
This suggests that the nonuniversal behavior of $P(\ln T)$ is associated with structural correlation.

\section{Measurement of optical transmission}

In the previous sections, we have considered simulations in samples with layers of random but uniform thickness. In practice, however, the thickness of each of the layers  is not perfectly uniform~\cite{Zhang}. Here we consider the applicability of the results above to a medium with nonuniform layers. 
We have measured the transmission of a helium-neon laser at 632.8 nm incident normally upon stacks of
22 mm$^2$ glass slides with an index of refraction $n = 1.5217$ and random thicknesses in the range 125-135 $\mu$m. Each slide has a roughly uniform fringe spacing, which varies over a wide range between 160 to 6800 $\mu$m. This gives a variation in wedge angle for individual slides of between $1.5{\mathrm{x}}10^{-5}$ and $2.6{\mathrm{x}10^{-3}}$ rad ~\cite{Zhang}. 
The air spacings between the layers are also random due to the nonuniformity in the glass slides and dirt particles between the
glass surfaces 

\begin{figure}
\centering
\includegraphics[width=0.8\columnwidth]{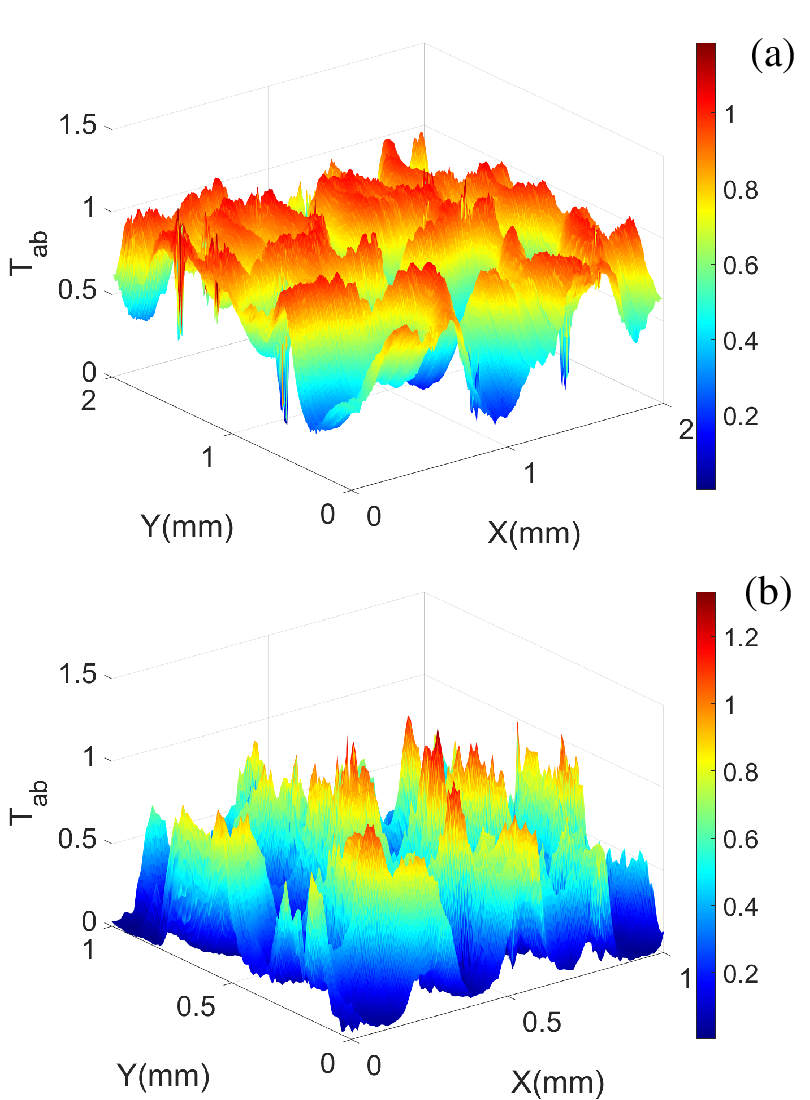}
\caption{Intensity pattern in transmission for a single sample for N=5 (a) and N=20 (b).}
\label{fig_9_sub}
\end{figure}

The single-frequency helium-neon laser was lightly focused on the layers of glass. A single-mode optical fiber is placed just behind the sample. Measurements of
transmission were made by scanning the sample in the plane of the glass slides
over a $2\times 2$ mm$^2$ area on a 10 $\mu$m grid for each sample. Measurements
were repeated for 10 and 6 different disordered configurations for the samples with $N=5$ and $N=20$, respectively. As a result of the nonuniform thickness of the layers, the disorder is not one-dimensional, and the transmitted intensity is highly nonuniform, as seen in Figs.~\ref{fig_9_sub}(a) and 9(b) for samples with $N=5$ and $N=20$, respectively. 

\begin{figure}
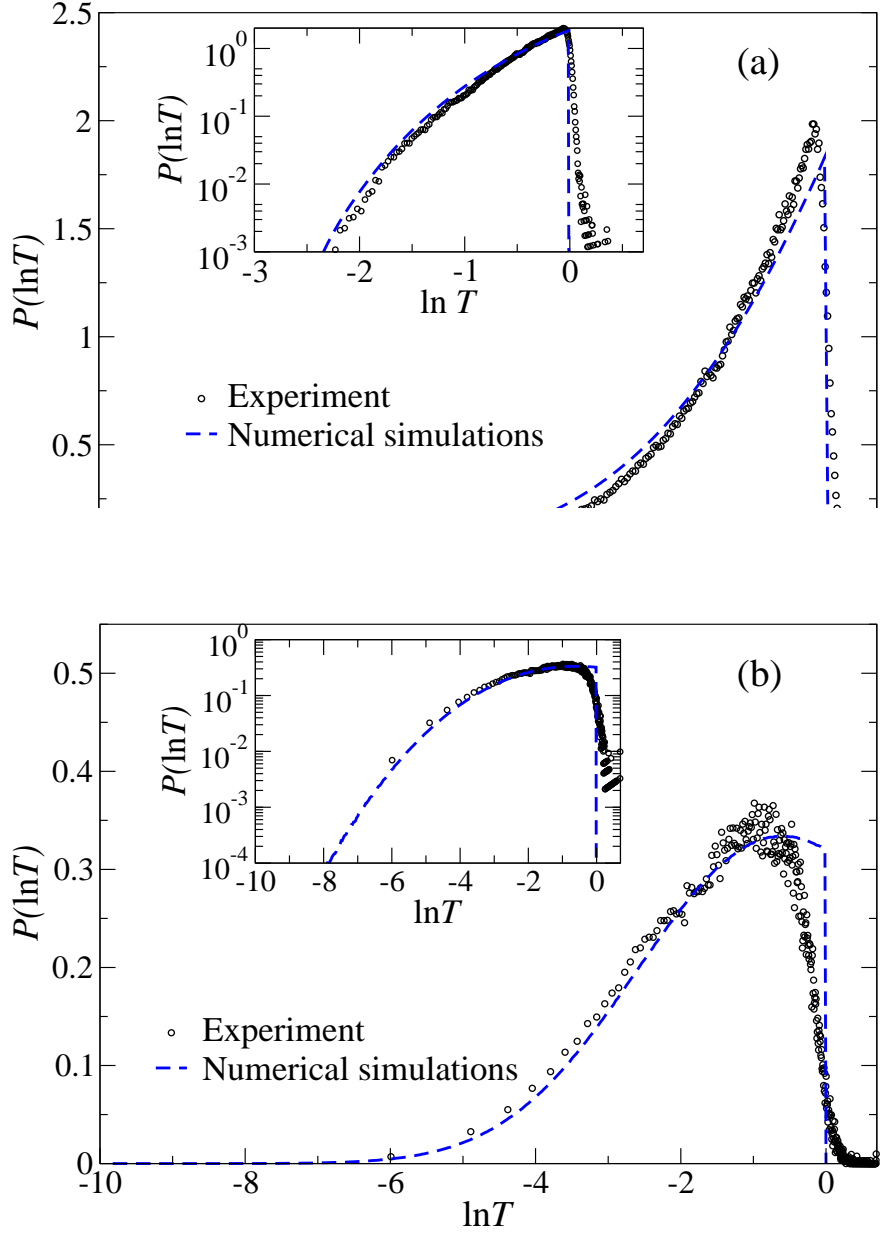

\centering
\includegraphics[width=0.8\columnwidth]{Fig_10a_RE_resub.eps}
\includegraphics[width=0.8\columnwidth]{Fig_10b_RE_resub.eps}
\caption{Optical measurements of $P(\ln T)$ (circles) for a glass stacks with $N=5$ (a) and $N=20$ (b) compared to the results of simulations represented by dashed lines, in Figs. 1(b) and 1(c), respectively. The inset shows 
$P(\ln T)$ on a logarithmic scale (y-axis). The distributions $P(\ln T)$ are obtained from  
$4\times10^5$ ($6\times10^4$) realizations of disorder for $N=5$ ($N=20$).}
\label{fig_10_sub}
\end{figure}

The distribution $P(\ln T)$ found in experiments for $N=5$ and $N=20$ (black circles), are shown in
Figs. \ref{fig_10_sub}(a) and \ref{fig_10_sub}(b), respectively. 
The bulk of the distributions are in reasonable agreement with simulations
for random binary systems, though substantial differences appear near $\ln T=0$. where the intensity can exceed the incident intensity. Whereas the transmitted intensity cannot exceed unity in a 1D medium, the intensity may be enhanced at points in the output for a layered sample that is not strictly one-dimensional as a result of transverse inhomogeneity in the thicknesses of the glass and air layers.  The intensity is therefore not uniform across the output and energy conservation does not require that the intensity at points at the output be less than unity~\cite{Zhang}.

Despite the strong variation of intensity in stacks of nonuniform slides, $P(\ln T)$ for the two distributions are seen in Fig.~\ref{fig_10_sub} to be similar to 1D simulations except at values of transmission near unity. At points of high transmission, the transmission time of light within the sample is large and the wave explores a greater range of nonuniformity, so large deviations from 1D simulations are observed.

\section{Conclusion}

We have considered the pdf of the logarithm of the transmission for random layered media, in which the phase shift modulo $\pi$ accumulated between layers is fully randomized. We find that $P(\ln T)$ has a universal segment above a crossover in transmission, corresponding to the result of RMT in the dense-weak-scattering limit of Eq.~(\ref{poflnT_exact}), and a nonuniversal segment determined by Eqs.~(\ref{Tm}) and (\ref{proptoalpha}). Whereas the statistics above the crossover between these segments depend upon a single parameter, $s=L/\ell$, the statistics below the crossover depend on two parameters. The parameters may be either $s$ and the number of bilayers $N$, or the reflectivity at the interface and $N$. The details of the structure have the greatest impact upon transmission statistics for thin samples with $s<1$. The transmission has a distinct minimum value $T_m$ equal to the transmission in the center of the band gap of the corresponding
quarter-wave stack~\cite{Freilikher}. 
As the number of layers increases for fixed $s$, the minimum transmission $T_m$ falls towards zero and $P(\ln T)$ evolves
towards the universal distribution given by Eq.~(\ref{poflnT_exact}) ~\cite{Melnikov,Abrikosov,Mello-book}.

The statistics of time delay  and energy density within the sample show that the universal and nonuniversal segments of $P(\ln T)$ are associated with longer and shorter transmission times, respectively, corresponding to higher and lower excitation of energy within the medium. Configurations falling in the nonuniversal segment of $P(\ln T)$ are more strongly correlated with a quarter wave stack, which gives the shortest transmission time and the most rapid drop of energy density within the medium, as well as the lowest transmission. 
Because the transmission time in configurations with low transmission is short, the spatial spread of the wave in such samples is limited and the nonuniformity of the layer thickness does not have a large impact on the pdf of transmission. The strongest deviations are associated with values of
intensity that exceed the incident level. 
The strongest deviations are associated with values of intensity that exceed the incident level. In these cases, the transmission time is large, and the transverse spread of the wave is large. The waves then traverse the same layer at different points. Transmission is then no longer uniform across the sample output, and fields may interfere to produce local fluctuations in intensity across the output with values that are greater than the incident intensity. 

In random layered systems with small fluctuations in layer thickness, the stop band would not be washed out, as it is for large fluctuations in layer thickness considered here. The pdf of $P(\ln T)$ would then depend upon frequency and would be broadest near the band edge~\cite{Faggiani}.

\section*{Acknowledgements}

This work is supported by the National Science Foundation (US) under EAGER Award No. 2022629
and under NSF-BSF Award No. 2211646.
V.A.G. acknowledges the financial support by MCIU (Spain) under the
Project number PID2022-136374NB-C22.

\appendix

\section{Transmission statistics near the lower cutoff}

Summing the amplitudes of all transmitted rays in a single layer with index $n$ and thickness $d$, the transmission is given by \cite{Fowles}:
\begin{equation}
\label{T_single}
T=\frac{{T_1}^2}{1+{R_1}^2-2{R_1}\cos\theta}
\end{equation}
where, $\theta=2nkd$, is the phase difference accumulated in a double pass through the layer, $k$ is
the wave vector and $R_1$ and $T_1$ are the reflection and transmission coefficients, respectively,
given by $T_1=4n/(n+1)^2$, and $R_1=(n-1)^2/(n+1)^2$. Since the random variation in thickness is
much greater than the wavelength, $\theta$ is assumed to be uniformly distributed. Thus
\begin{eqnarray}
P(\ln (T-T_m))&=&p(\theta)\left| \frac{d\theta}{d\ln(T-T_m)} \right| \nonumber \\
&& = \frac{1}{2\pi}(T-T_m)\left| \frac{d\theta}{dT}\right| ,
\label{poflnTminusTm}
\end{eqnarray}
with $p(\theta)=1/2\pi$. The derivative $d\theta/dT$ in Eq.~(\ref{poflnTminusTm}) can be found from Eq.~(\ref{T_single}). This gives
\begin{equation}
 \cos \theta =\frac{1}{2R_1}\left(R_1^2+1-\frac{T_1^2}{T} \right)=\frac{1}{2R_1}\left(\frac{T_1^2}{T_m} -2R_1-\frac{T_1}{T} \right),
\end{equation}
where we have used that $T_m=T_1^2/(1+R_1)^2$. Taking the derivative respect to $\theta$, we have 
\begin{equation}
\frac{d\theta}{dT} =\frac{T_1 T_m}{T\sqrt{\left(T-T_m\right)\left(T_1^2\left(T-T_m \right)+4R_1T_m \right)}}
\end{equation}
For $T \to T_m$, we thus have 
\begin{equation}
\frac{d\theta}{dT} \approx \frac{1}{2}\frac{T_1}{\sqrt{R_1T_m}}\frac{1}{\sqrt{T-T_m}}
\end{equation}
Introducing this expression into Eq.~(\ref{poflnTminusTm}), we obtain
\begin{equation}
P\left(\ln(T- T_m)\right) \sim \frac{1}{4\pi} \frac{T_1}{\sqrt{R_1 T_m}}\sqrt{T-T_m} .
\label{poflnTminusTmapprox}
\end{equation}
This corresponds to  $\alpha=1/2$ for a single layer independent of the refractive index, in accord with 
simulations. 

For a single layer, $N=1$, there is a single element with random thickness and random
phase between configurations. For every additional binary layer of the random sample there are an
additional two spacings in which a random phase is accumulated in propagation through the element. 
Since the disorder in different layers is independent, this suggests that the total exponent of
$(T-T_m)$ in $P\left( \ln(T-T_m)\right)$ is proportional to the number of spacings with random thickness in the sample. This leads to Eq.~(\ref{alpha}) in the main text, which is in accord with the results shown in Fig.~\ref{fig_3_sub} for $N\le 4$, in the main text.

The complete pdf of $\ln T$ for $N=1$ can readily be found by averaging
Eq.~(\ref{T_single}) over $\theta$. This gives
\begin{equation}
\label{poflnT_1}
P(\ln T) = \frac{T_1^2}{\pi T \sqrt{4R_1^2-\left(1+R_1^2 -T_1^2/ T \right)^2}} . 
\end{equation}
This result is plotted in the inset Fig.~\ref{fig_2_sub}(a) in the main text and coincides with the results of simulations.

\end{document}